\documentclass[10pt]{article}
\usepackage[OE]{express}

\begin{document}

\title{Pump induced lasing suppression in Yb:Er-doped microlasers}

\author{Fuchuan Lei,\authormark{1*}, Yong Yang, Jonathan M. Ward and S\'ile Nic Chormaic}
\address{\authormark{1}Light-Matter Interactions Unit, Okinawa Institute of Science and Technology Graduate University,
Onna, Okinawa 904-0495, Japan. \\
}
\email{\authormark{*}fuchuan.lei@oist.jp} 



\begin{abstract}
A pump source is one of the essential prerequisites in order to achieve lasing, and, in most cases, a stronger pump leads to higher laser power at the output. However, this behavior may be suppressed if two pump beams are used.  In this work, it is shown that lasing around the 1600 nm band can be suppressed completely if two pumps, at wavelengths of 980 nm and 1550 nm, are applied simultaneously to a Yb:Er-doped microlaser, whereas it can be revived by switching one of them off. This phenomenon can be explained by assuming that the existence of one pump (980 nm) changes the role of the other pump (1550 nm); more specifically, the 1550 nm pump starts to consume the population inversion instead of increasing it when the 980 nm pump power exceeds a certain value. As a result, the two pump fields lead to a closed-loop transition within the gain medium (i.e., erbium ions). This study unveils an interplay similar with coherence effects between different pump pathways, which provides a reference for designing the laser pump and may have  applications in lasing control. 
\end{abstract}



\section{Introduction}
The presence of a pump can be considered to be one of the most fundamental elements to achieve lasing in a system. In solid state lasers, the function of the pump is to invert the population by transferring ions from ground states to metastable states via intermediate states with the assistance of a quick decay channel. For many gain media, there are several intermediate states from which ions can decay to the same metastable state \cite{koechner2013solid}. This implies that the laser can be pumped at several wavelength bands separately or co-operatively. For instance, erbium ions have two frequently-used pump bands to produce emission in the 1.55 $\mu$m region, i.e., 980 nm and 1450-1480 nm, corresponding to  transitions from  the ground-state manifold $\rm ^4I_{15/2}$ to the $\rm ^4I_{11/2}$ and $\rm ^4I_{13/2}$ manifolds, respectively \cite{becker1999erbium}. Since the absorption cross-section is high at 980 nm and the absorption band is wide at 1450 nm, it is quite natural to combine the two pumps in order to gain higher laser power output in the C-band \cite{sini,Bray99,koch1997optical}. Aside from providing gain in the C-band, the Er$^{3+}$ ions pumped by 1530-1550 nm light may also be used as a gain medium for L-band lasing \cite{choi2001high,lee1999enhancement,young2004efficient}.

In this work, we experimentally observe that, unlike the 980 nm and 1480 nm co-pumping case, L-band lasing can be enhanced or completely \textit{suppressed} when two pumps at 980 nm and 1550 nm are simultaneously applied to a Yb:Er co-doped laser. This counter-intuitive phenomenon indicates that it is incorrect to simply sum up the effects induced by the two pumps, since they may actually have opposing effects and lead to a reduction in the lasing output. According to our experimental results and theoretical model, we reveal that the lasing suppression phenomenon originates from the emergence of a closed transition cycle for the gain medium formed by the two pumps. Therefore, this phenomenon should not be limited exclusively to Yb:Er lasers or even solid state lasers - it may well exist in other laser systems that are co-pumped at different wavelengths. One would thus expect this phenomenon to have striking consequences for laser pumping or control.

\section{Experimental Method}
Here, we will consider a variety of lasers with different gain media and ion concentrations.  For convenience, we use silica microsphere lasers with whispering gallery (WGM) cavity modes.  In the last decades, lasers based on WGMs have attracted significant attention for both fundamental research and applications \cite{reynolds2017fluorescent,he2013whispering,he2011detecting,hodaei2014parity,feng2014single,peng2014loss,yang2016tunable,ward2016glass,ward2008taper,wu2010ultralow,peng2016chiral}. We make full use of the advantages of WGM lasers, such as ease off fabrication and low cost. The diameters of our microspheres are about 150 $\mu$m and they are fabricated in a standard way by focusing a $\rm CO_{2}$ laser beam directly onto a piece of commercial fiber (Thorlabs, SMF 28). Initially, a section of fiber is mounted vertically and  tapered by heating via the $\rm CO_{2}$ laser while a small weight pulls it downwards.  The fiber is next cut using the $\rm CO_{2}$ laser beam and the tapered part acts as the stem of the microsphere. This thin section of fiber is dipped for a few seconds into a sol-gel precursor solution in which erbium and ytterbium ions have been dissolved. The sol-gel precursor fabrication method has been discussed in detail elsewhere \cite{yang2005erbium,yang2016tunable}. The sphere is then formed at the sol-gel coated section via a reflow process using the $\rm CO_{2}$ laser. During the reflow process, the residual sol-gel solvent is removed and the ions are embedded into the silica matrix due to the high temperature. This fabrication method means that the precise concentration of the erbium and ytterbium ions cannot be obtained; however, their ratio and  maximum concentrations can be determined by the sol-gel precursor. Hence, concentrations stated in the following are actually the maximum possible values.

To pump the microsphere laser and couple laser emissions out from it, a tapered optical fiber with a diameter of 1-2 $\mu$m is used in a contact coupling configuration. We use three broadband, multiple-mode lasers, with 5-10 nm linewidths and  wavelengths centered at 980 nm, 1440nm, and 1550 nm, as pump sources. Though the frequency of the pump lasers cannot be tuned, some light can couple into the microsphere with  a few percent of the light being absorbed by the cavity modes. For co-pumping of the microsphere by two lasers simultaneously, we combine the two beams using wavelength-division multiplexing (WDM) before coupling them into the tapered fiber.  To avoid possible damage, only 50\% of the laser output \textbf{is} sent into an optical spectrum analyzer (OSA) for measurement, the resolution of which is set to 0.5 nm. 

\section{Experimental Results}
\begin{figure}[!b]
  \centering
  \includegraphics[width=4in]{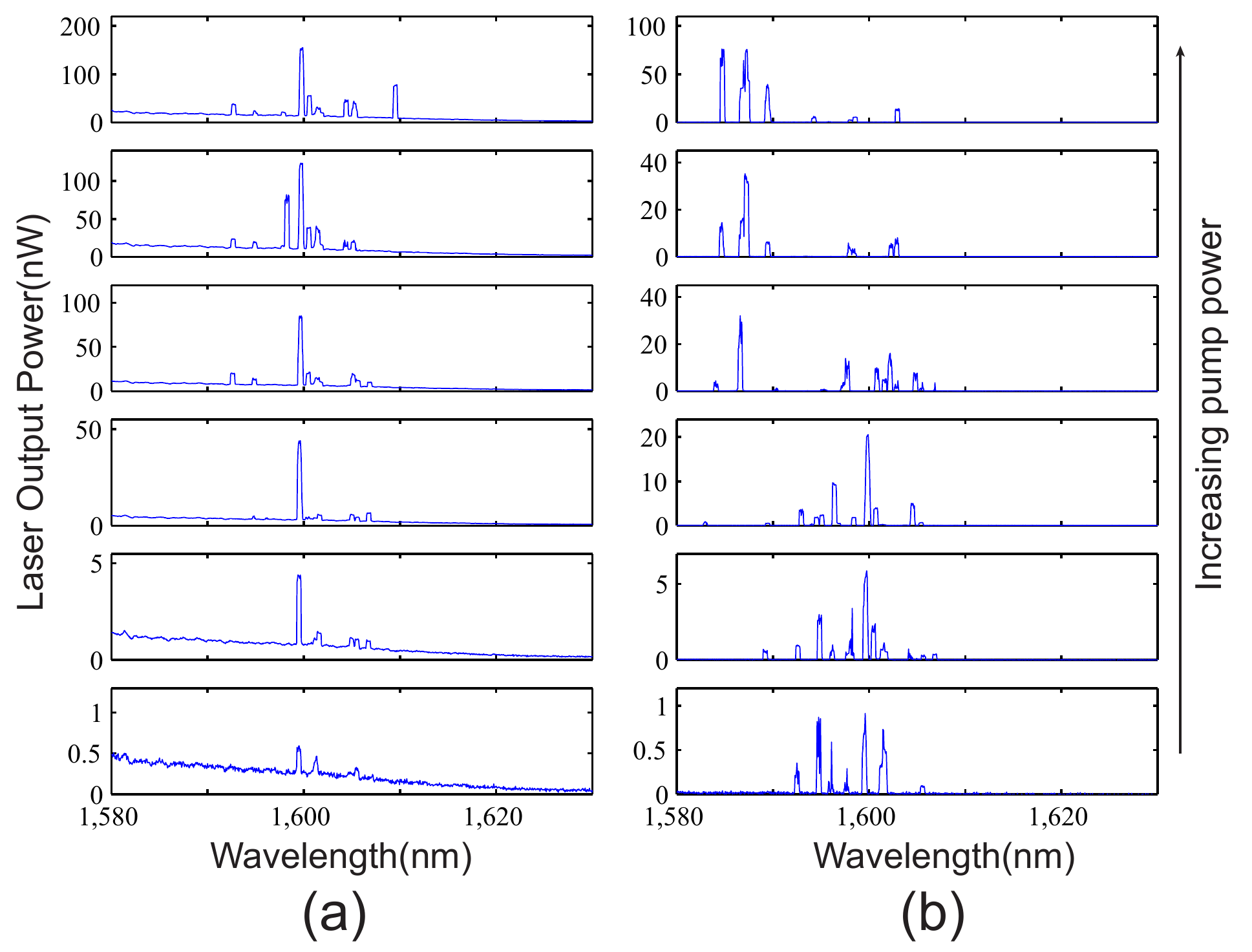}
  \caption{Output spectra of the Yb:Er codoped microsphere laser pumped by (a) a 1550 nm laser and (b) a 980 nm laser. From bottom to top the power of the 1550 nm laser launched into the coupling fiber is 0.8, 2.1, 7.7, 17.5 ,27, 35.4 mW and the power of the 980 nm laser is 1, 6.9, 13.6, 34.8, 65.1, and 72.2 mW.  The 980 nm pump emissions tend to shift towards the blue for higher pump powers. }
  \label{fig1}
  \end{figure}
  
  \begin{figure}[!ht]
  \centering
  \includegraphics[width=4.5in]{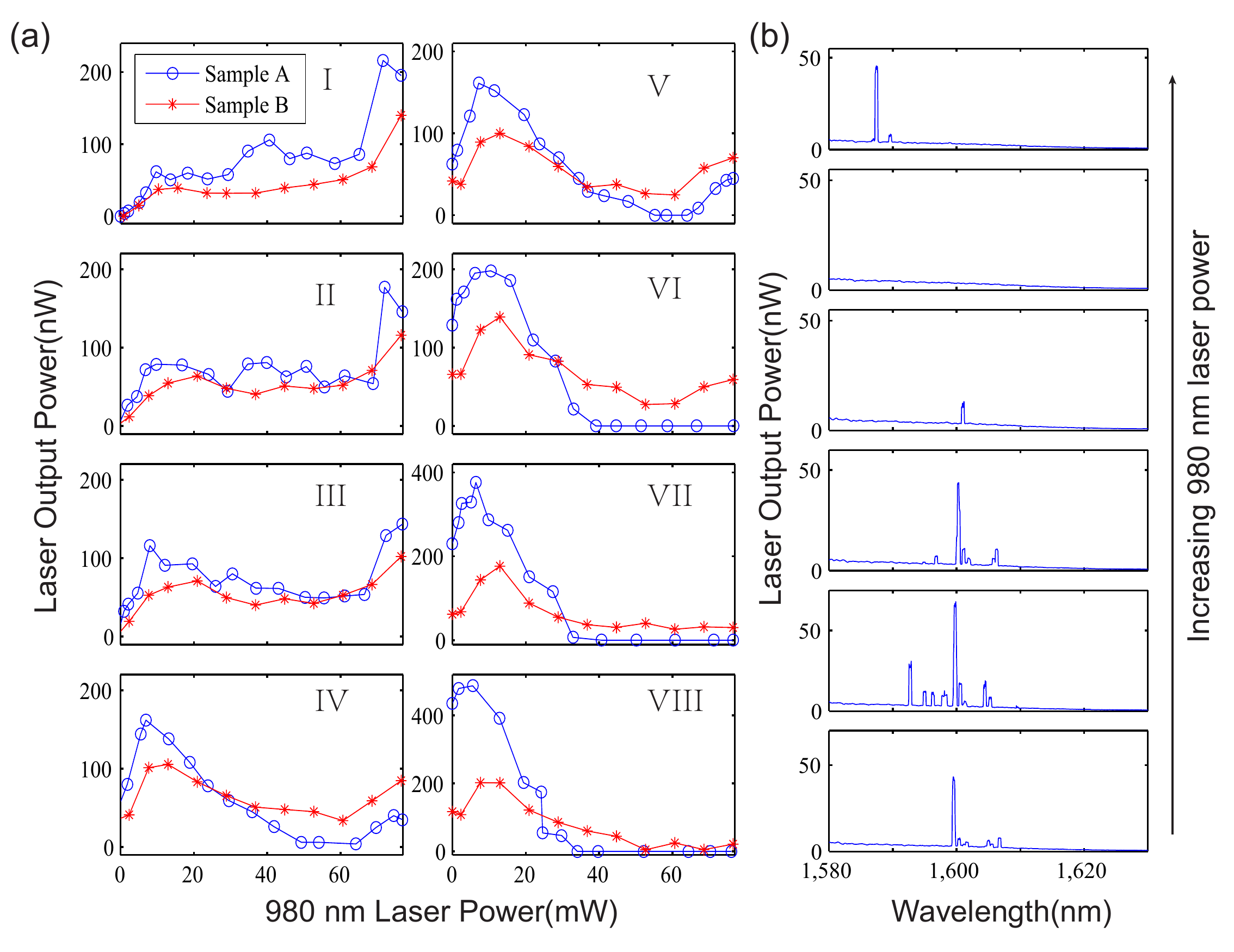}
  \caption{Output of the Yb:Er co-doped microsphere laser pumped by the 1550 nm laser and the 980 nm laser simultaneously. (a) Total output power as the 980 nm pump is increased from 0 to 76.7 mW. For convenience, the total output power of sample B is amplified 3 times. From $\rm \uppercase\expandafter{\romannumeral1}$ to $\rm \uppercase\expandafter{\romannumeral8}$, the 1550 nm power launched into the coupling fiber is 0, 1.4, 3.2, 6.8, 7.9, 14, 21.7, and 34.9 mW for sample A and 0, 1.2, 2.4, 7.2, 9.6, 12, 14.4, and 18.2 mW for sample B. (b) Output spectra of sample A as the 980 nm pump power is increased. From bottom to top, the 980 nm power is 0, 7.2, 23.8, 36.9, 58.4, and 71.7 mW and the 1550 nm power is 7.9 mW, corresponding to (a) Plot $\rm \uppercase\expandafter{\romannumeral5}$. }
  \label{fig2}
  \end{figure}
First, we characterized the Yb:Er laser while it was pumped  by the 980 nm laser and the 1550 nm laser separately. The concentrations of the trivalent erbium and ytterbium ions in the surface layer of the microspheres were $\rm 5\times10^{19}\slash cm^{3}$ and $\rm 2\times10^{20}\slash cm^{3}$, respectively.
The WGM laser demonstrated multiple-mode output at the L-band (around 1600 nm, see Fig.\ref{fig1}). It was also observed random mode hopping; therefore, the spectra were sampled stochastically. For both pump wavelengths, the number of lasing modes and the total output power increased with  pump power. It was also noted, see Fig. \ref{fig1}(b), that some lasing peaks had a tendency to shift towards shorter wavelengths, i.e., towards 1580 nm, when the power of the 980 nm pump reached 13.6 mW. However, the shortest wavelengths are still longer than 1580 nm. This phenomenon can be attributed to the complex energy level structure of Er$^{3+}$ as discussed elsewhere \cite{cai2002laser,arnaud2004microsphere}.

Figure \ref{fig2} shows the output of the Yb:Er co-doped WGM laser pumped simultaneously by  the 1550 nm laser and 980 nm laser. The concentrations of the two samples are the same as in the above. Figure \ref{fig2}(a) shows the total power of the WGM laser output, calculated from the measured spectra on the OSA. For each plot, the power of the 1550 nm pump is fixed while the 980 nm pump launched into the coupling fiber is increased from 0 to 76.7 mW. From plots $\rm \uppercase\expandafter{\romannumeral1}$ to $\rm \uppercase\expandafter{\romannumeral8}$, the power of the 1550 nm laser is also increased. For clarity, the following discussion is based on one sample (A) only, since sample A and B are fabricated by using the same method and have similar sizes. As shown in plots $\rm \uppercase\expandafter{\romannumeral1}$ and $\rm \uppercase\expandafter{\romannumeral2}$ of Fig. \ref{fig2}(a), when the power of the 1550 nm laser is low, the total power of the WGM laser output increases with the 980 nm pump. When the power of the 1550 nm pump increases to a certain level (6.8mW), the total power of the laser output start to decrease after a slight increase, as shown in plots $\rm \uppercase\expandafter{\romannumeral3}$ to $\rm \uppercase\expandafter{\romannumeral5}$; however, it can revive after reaching  zero. As seen in plots $\rm \uppercase\expandafter{\romannumeral6}$,$\rm \uppercase\expandafter{\romannumeral7}$ and $\rm \uppercase\expandafter{\romannumeral8}$ of Fig. \ref{fig2}(a), when the power of the 1550 nm pump is high, the total power of the WGM laser may be completely suppressed after slight augmentation of the 980 nm pump. Figure \ref{fig2}(b) shows the observed WGM output spectra of sample A. From bottom to top, the 980 nm pump power has values of 0, 7.2, 23.8, 36.9, 58.4, and 71.7 mW, corresponding to Fig. \ref{fig2}(a) plot $\rm \uppercase\expandafter{\romannumeral5}$.

\begin{figure}[!ht]
 \centering
 \includegraphics[width=4.5in]{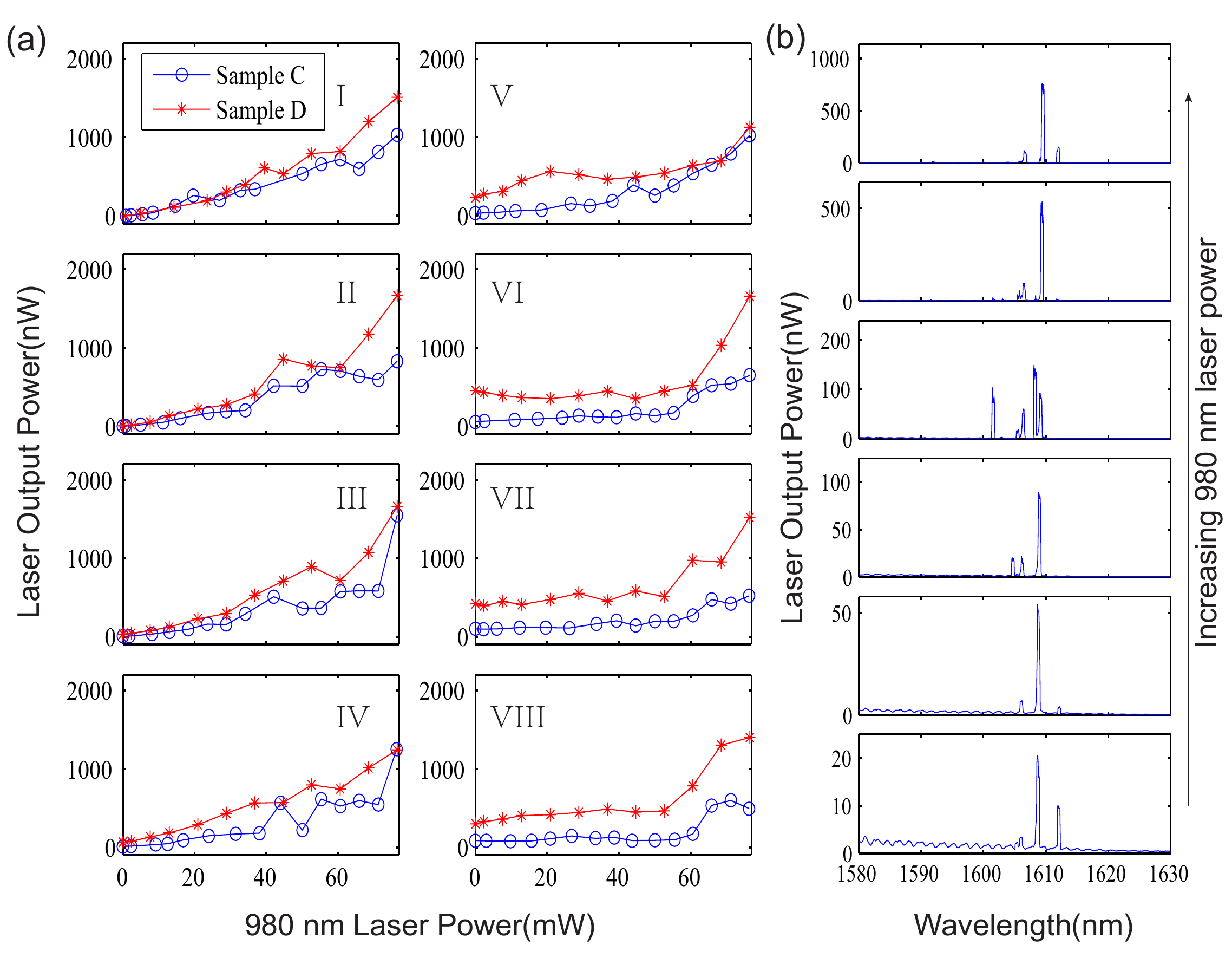}
 \caption{Output of the Er-doped microsphere laser pumped at 1550 nm and 980 nm simultaneously. (a) As in Fig. \ref{fig2}, the total output power of the WGM laser with respect to different pump powers. From  plot $\rm \uppercase\expandafter{\romannumeral1}$ to plot $\rm \uppercase\expandafter{\romannumeral8}$, from bottom to top, the 1550 nm power launched into the coupling fiber is 0, 1.7, 4.3, 8.65, 17.3, 34.6, 52 and 95 mW for sample C. For sample D, the 1550 nm power is 0, 1.2, 5.8, 11.5, 34.5, 57.5, 69 and 80.5 mW. Sample C and D are fabricated by using the same method and have similar sizes. (b) Output spectra of sample C when increasing the 980 nm pump power, corresponding to the blue (open circles) line in (a) plot$\rm \uppercase\expandafter{\romannumeral5}$. From bottom to top, the 980 nm power is to 0, 7.2, 23.8, 36.9, 58.4, 71.7 mW, and the 1550 nm power is 7.9 mW.}
 \label{fig3}
 \end{figure}
For comparison, we did similar experiments for microspheres doped purely with Er$^{3+}$ ions, see Fig. \ref{fig3} . The concentration of Er$^{3+}$  was kept at $\rm 5\times10^{19}\slash cm^{3}$.  As in Fig. \ref{fig2}, we increased the power of the 980 nm pump for each plot. From Plot $\rm \uppercase\expandafter{\romannumeral1}$ to Plot $\rm \uppercase\expandafter{\romannumeral8}$, the power of the 1550 nm pump was also increased gradually. In qualitative contrast to the observations for the co-doped WGM laser, we clearly see that the total output power keeps increasing as a function of the 980 nm power, no matter the power of the 1550 nm pump. Figure \ref{fig3}(b) depicts a series of output spectra,  corresponding to the blue data (open circles)  in plot $\rm \uppercase\expandafter{\romannumeral5}$.

\begin{figure}[!ht]
\centering
\includegraphics[width=4.5in]{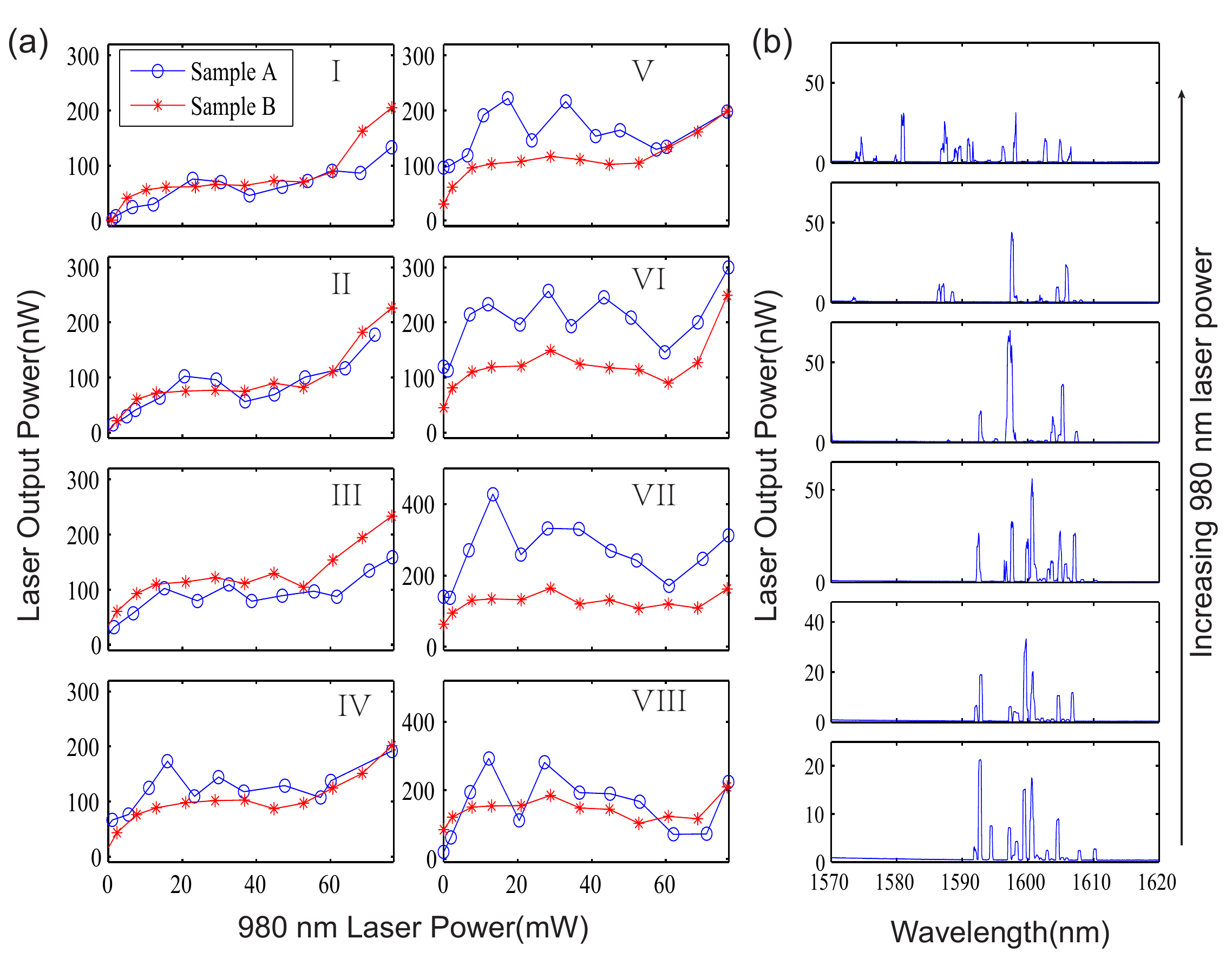}
\caption{Output of the Yb:Er co-doped microsphere laser pumped at 1440 nm and 980 nm simultaneously. (a) As in Fig.2, the total output power of the WGM laser with respect different pump powers. From $\rm \uppercase\expandafter{\romannumeral1}$ to $\rm \uppercase\expandafter{\romannumeral8}$, the 1440 nm pump power is 0, 2.3, 9.3, 30.3, 47.5, 62.8, 80 and 97.7 mW for sample A; For sample B, the 1440 nm power is 0, 7.5, 17.1, 26.1, 36.5, 47.7, 58.1 and 63.3 mW, respectively. For clarity, the total output power for sample B is amplified 5 times. (b) The output spectra of sample A when increasing the 980 nm pump power, corresponding to the blue (open circles) line in (a) plot $\rm \uppercase\expandafter{\romannumeral5}$.From bottom to top, the 980 nm power is 0, 6.4, 17.4, 32.9, 57.4, 76.5 mW, and 1440 nm power is 47.5 mW.}
\label{fig4}
\end{figure}
To further inspect the effect of the pump wavelength on WGM lasing, we conducted another group of experiments. The microsphere samples are the same two as used for Fig. \ref{fig2}, but we use a 1440 nm pump instead of the 1550  one, and the results are given in  Fig. \ref{fig4}. From plot $\rm \uppercase\expandafter{\romannumeral1}$ to plot $\rm \uppercase\expandafter{\romannumeral8}$, we increased the 1440 nm pump power  from 0 to 97.7 mW for sample A and from 0 to 63.3 mW for sample B. When the 1440 nm pump power is low (plot $\rm \uppercase\expandafter{\romannumeral1}$- plot $\rm \uppercase\expandafter{\romannumeral3}$) the output power of the WGM laser
increases with respect to the 980 nm pump. When the 1440 nm pump power is increased to 62.8 mW (plot $\rm \uppercase\expandafter{\romannumeral6}$- plot $\rm \uppercase\expandafter{\romannumeral8}$), the total output power of the laser initially increases and gradually reaches a maximum value, though with some fluctuations. In particular, sample A exhibits considerable power fluctuations. Since the pump laser we used supports multiple-mode emission, and considering mode hopping, it is impossible to conclude whether the lasing is suppressed or enhanced by the 980 nm pump. 
\begin{figure} [h]
\centering
\includegraphics[width=4.3in]{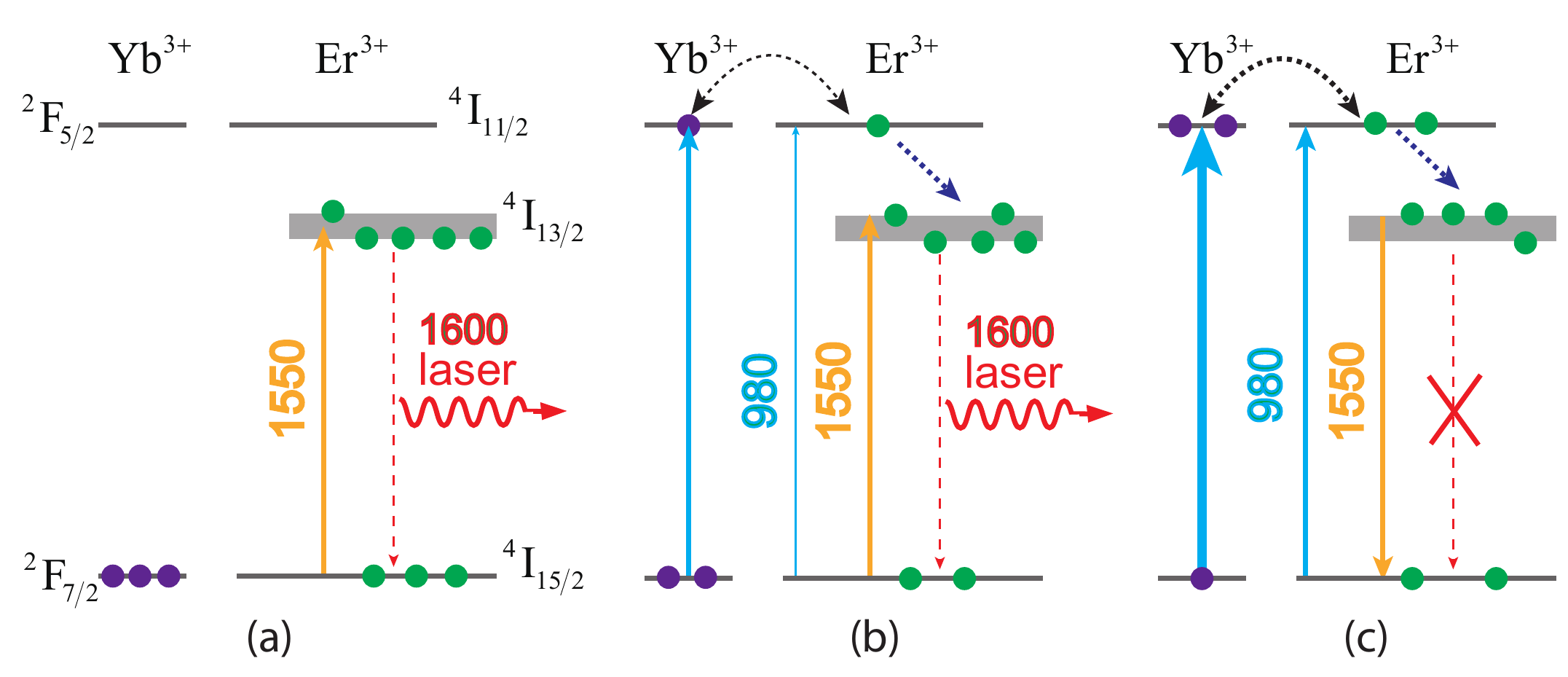}
 \caption{Illustration of pump induced lasing suppression in a Yb:Er co-doped microlaser. (a) Pumping via a strong 1550 nm laser; (b) Co-pumping via a strong 1550 nm laser and a  weak 980 nm laser; (c) Co-pumping via a strong 1550 nm laser and a strong 980 nm laser.}
 \label{fig5}
\end{figure}

\section{Theoretical Model and Discussion}

To better understand the experimental results, we  consider the energy levels of the Er$^{3+}$ and Yb$^{3+}$ ions, see  Fig.\ref{fig5}. It is well-known \cite{white2007resonant} that the 1600 nm lasing emission is generated by the transition from the $\rm ^4I_{13/2}$ to $\rm ^4I_{15/2}$ energy level in Er$^{3+}$. Due to  Stark splitting effects \cite{becker1999erbium}, the energy levels of rare earth ions in an SiO$_2$  matrix are in fact energy manifolds consisting of many sublevels, and, as a result, the transitions corresponding to pumping by 1440 nm  and 1550 nm, as well as the 1600 nm emission, occur between the same two manifolds, i.e., $\rm ^4I_{13/2}$ and $\rm ^4I_{15/2}$. Though the energy level structures of rare earth ions are very complicated, some phenomenological parameters, such as absorption and emission cross-sections, can be introduced to encompass the main physical properties. For example, with the assumption that 1550 nm light has a higher absorption to emission cross-section ratio than 1600 nm light, we can illustrate why the former can be used as a pump to produce the latter. 

In the Yb:Er co-doped laser, the 980 nm pump can excite both  Er$^{3+}$ and  Yb$^{3+}$  \cite{ward2007optical, paschotta1997ytterbium}, corresponding to the transitions from their ground states to the levels $\rm ^4I_{11/2}$ and  $\rm ^2F_{5/2}$, respectively, see Fig. \ref{fig5}(a). Compared to Er$^{3+}$,  Yb$^{3+}$  can absorb the 980 nm pump more efficiently and energy can be transferred to the erbium ions in the ground-state manifold. In principle, Er$^{3+}$ can also transfer energy back to  Yb$^{3+}$; however, the probability of this happening is negligible due to the short lifetime (less than 1 $\mu$s\cite{wu2003fluorescence}) of the Er$^{3+}$ intermediate state, $\rm ^4I_{13/2}$ \cite{paschotta,da2003mechanism}. In our system, the concentration of Yb$^{3+}$   is higher than that of  Er$^{3+}$, therefore, the effective absorption and emission cross-sections for the Yb$^{3+}$-sensitized Er$^{3+}$  ions at the 980 nm band are significantly increased. For simplicity and clarity, in the following we will  ignore the complicated Yb:Er energy exchange dynamics and simply treat the co-doped medium as a three-level atom, with an energy level structure similar to that of Er$^{3+}$. The differences we consider are only the absorption and emission cross-sections at the 980 nm band. In our model, we assume that the former is an order of magnitude larger than the latter. 

As a proof-of-principle, we assume that the cavity is pumped by two single mode lasers, one   at 980 nm and the other at either 1440 nm or 1550 nm. For the waveguide-coupled WGM cavity, the equations describing the pump and  laser fields in the cavity are \cite{lei2014dynamic}
\begin{eqnarray}
\frac{da_{p_1}}{dt} &=& -\left(\frac{\kappa^{0}_{p_{1}}+\kappa^{ex}_{p_{1}}-g_{p_1}}{2}\right)a_{p_1}-\sqrt{\kappa^{ex}_{p_{1}}}a_{p_{1}}^{in},\label{eq1}\\
\frac{da_{p_2}}{dt} &=& -\left(\frac{\kappa^{0}_{p_{2}}+\kappa^{ex}_{p_{2}}-g_{p_2}}{2}\right)a_{p_2}-\sqrt{\kappa^{ex}_{p_{2}}}a_{p_{2}}^{in},\label{eq2}\\
\frac{da_{L}}{dt} &=& -\left(\frac{\kappa^{0}_{p_{L}}+\kappa^{ex}_{p_{L}}-g_{L}}{2}\right)a_{L}+\xi,\label{eq3}
\end{eqnarray}
where the subscripts $p_1$, $p_2$ and $L$ denote the 980 nm pump, the 1440 or 1550 nm pump, and the 1600 nm lasing, respectively; $a$ is the amplitude of the intracavity field, $\kappa^0$ and $\kappa^{ex}$ are the decay ratios caused by intrinsic losses and fiber coupling loss, respectively; $a^{in}$ designates the input field in the fiber and $|a^{in}|^2$ is the input power. For simplicity, we also assume that each input field is resonant with its corresponding cavity modes. $\xi$ is  the input noise to the laser mode originating from spontaneous emission; $g_{p_1}$, $g_{p_2}$ and $g_L$ denote the gain coefficients for the 980 nm pump, the 1440/1550 nm pump, and the 1600 nm lasing, respectively, and are determined from
\begin{eqnarray}
g_{p_{1}} &=& \frac{c}{n_{p_{1}}}\left(N_{3}\sigma_{p_1}^{e} - N_{1}\sigma_{p_1}^{a}\right),\label{eq4}\\
g_{p_{2}} &=& \frac{c}{n_{p_{2}}}\left(N_{2}\sigma_{p_2}^{e} - N_{1}\sigma_{p_2}^{a}\right),\label{eq5}\\
g_{p_{L}} &=& \frac{c}{n_{L}}\left(N_{2}\sigma_{L}^{e} - N_{1}\sigma_{L}^{a}\right),\label{eq6}
\end{eqnarray}
where $c$ denotes the speed of light in vacuum, $n$ stands for the effective refractive index, $\sigma^a$ and $\sigma^e$ describe the absorption and emission cross-sections of Er$^{3+}$ ,
 $N_1$, $N_2$ and $N_3$ are the density of Er$^{3+}$  in the states $\rm ^4I_{15/2}$, $\rm ^4I_{13/2}$, and  $\rm ^4I_{11/2}$, respectively. The dynamical behavior of this three-level system is described by the following laser rate equations
\begin{eqnarray}
\frac{dN_{3}}{dt} &=& -\frac{N_{3}}{\tau_{32}} +C_{p_{1}} \left(N_{1}\sigma_{p_{1}}^{a} - N_{3}\sigma_{p_{1}}^{e}\right)\left|a_{p_1}\right|^2,\label{eq7}\\
\frac{dN_{2}}{dt} &=& \frac{N_{3}}{\tau_{32}}-\frac{N_{2}}{\tau_{21}}+C_{p_{2}} \left(N_{1}\sigma_{p_{2}}^{a} - N_{2}\sigma_{p_{2}}^{e}\right)\left|a_{p_2}\right|^2 + C_{L} \left(N_{1}\sigma_{L}^{a} - N_{2}\sigma_{L}^{e}\right)\left|a_{L}\right|^2, \label{eq8}
\end{eqnarray}
where $\tau_{32}$ and $\tau_{21}$ denote the decay time for spontaneous emission from $\rm ^4I_{11/2}$ to  $\rm ^4I_{13/2}$ and from $\rm ^4I_{13/2}$ to
the ground level, $\rm ^4I_{15/2}$, respectively. $C_{p_1,p_2,L}=\lambda_{p_1,p_2,L}/{2\pi RA_{p_1,p_2,L}n_{p_1,p_2,L}h}$, where $A$ is the effective mode area for the probe and pump modes, $R$ is the radius of the microsphere and $h$ is Planck's constant.

Eqs. (\ref{eq1})-(\ref{eq8}) can be solved numerically. Here, we only consider the steady-state \textbf{case}. The values of the parameters we use in this model are as follows\cite{lei2014dynamic}: the density of Er$^{3+}$  is $\rm 2\times10^{19}/cm^{3}$, $R=75$ $\mu$m, $A_{p_1,p_2,L}=5\times10^{-8} \rm{cm^2}$, $\lambda_{p_1}=980$ nm, $\lambda_{p_2}=1440/1550$ nm, $\lambda_{L}=1600$ nm, $n_{p_1,p_2,L}=1.35$, $\tau_{21}=6$ ms, $\tau_{32}=1$ $\mu$s, $\sigma_{p_1}^a=3.3\times10^{-21} \rm{cm^2}$, $\sigma_{p_1}^e=2\times10^{-21} \rm{cm^2}$, $\sigma_{L}^a=3.3\times10^{-22} \rm{cm^2}$, and $\sigma_{L}^e=1\times10^{-21} \rm{cm^2}$. For the 1440 nm pump, $\sigma_{p_2}^a=2\times10^{-22} \rm{cm^2}$ and $\sigma_{p_2}^e=5\times10^{-23} \rm{cm^2}$, whereas for the 1550 nm pump, $\sigma_{p_2}^a=2.5\times10^{-21} \rm{cm^2}$ and $\sigma_{p_2}^e=3\times10^{-21} \rm{cm^2}$. For the Yb:Er co-doped laser, we use $\sigma_{p_1}^a=3.3\times10^{-20} \rm{cm^2}$ and $\sigma_{p_1}^e=2\times10^{-20} \rm{cm^2}$, as mentioned previously.

\begin{figure}[t]
  \centering
  \includegraphics[width=4.5in]{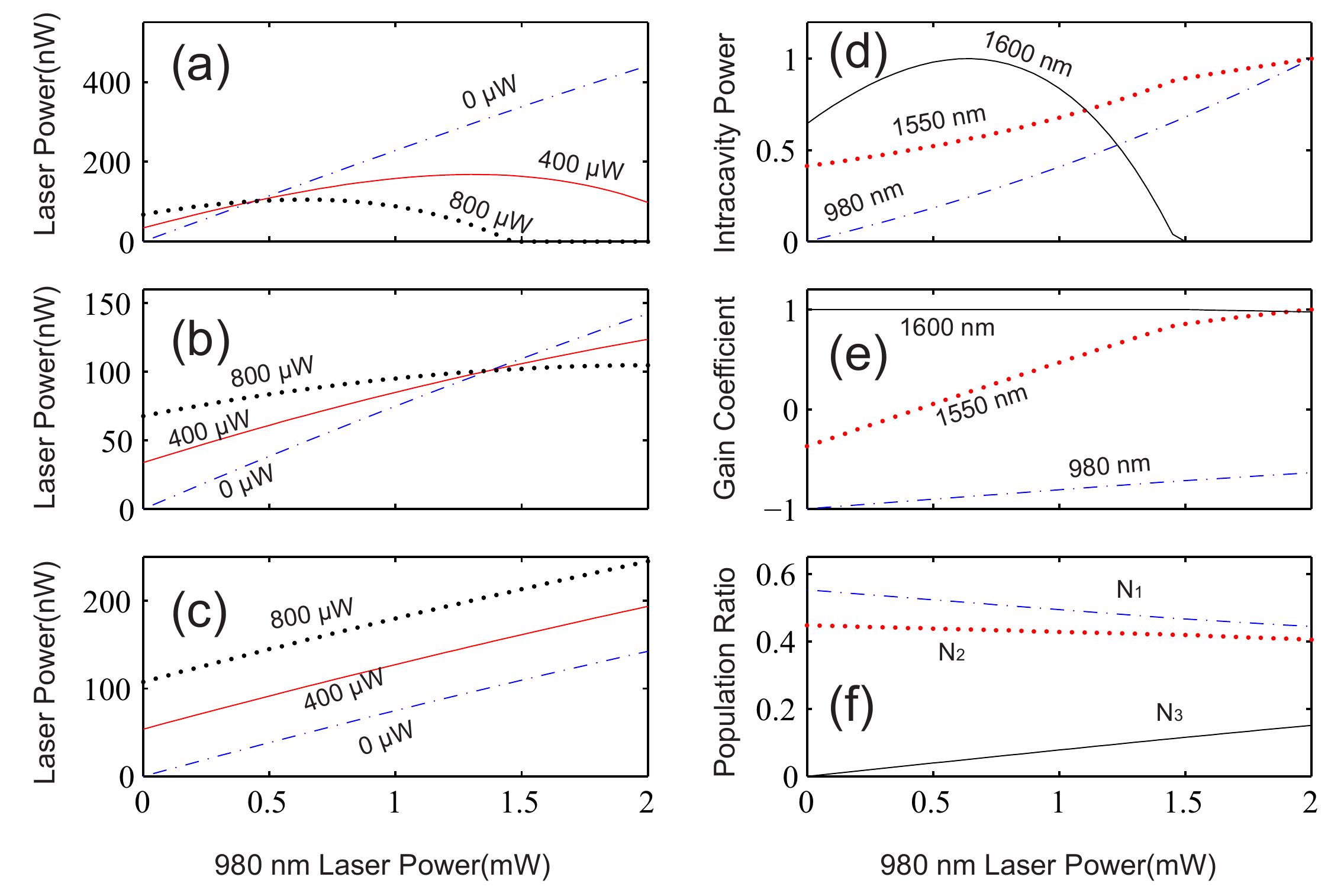}
   \caption{Results from the theoretical model. (a) and (b) show the output powers of Yb:Er co-doped and pure Er laser pumped by 980 nm and 1550 nm lasers, respectively. The three lines stand for different 1550 nm pump 
powers: 0 (blue dashed), 400 $\mu$W (red solid) and 800 $\mu$W (black dot). (c) shows the output powers of Yb:Er codoped laser pumped by 980 nm and 1440 nm lasers. The power of 1440 nm pump is set at 0 (blue dashed), 400 $\mu$W (red solid) and 800 $\mu$W (black dot), respectively. (d)-(f) depict some parameters of the Yb:Er codoped laser evolve with the 980 nm pump, and the 1550 nm  pump power is fixed at 800 $\mu$W. (d) and (e) stand for the intracavity power and gain coefficient of 980 nm pump (blue dashed), 1550 nm pump (red dot) and 1600 nm laser (black solid), respectively. (f) shows the population ratios of erbium ions: $N_1$: blue dashed, $N_2$: red dot, and $N_3$: black solid.}
   \label{fig6}
  \end{figure}

Figure \ref{fig6}(a)-(c) presents the theoretical estimates for the output power of the 1600 nm laser,  corresponding to the experiments shown in Figs. \ref{fig2}, \ref{fig3} and \ref{fig4}, respectively. From Fig. \ref{fig6}(a), the 1600 nm laser output increases with the 980 nm pump when the power of the 1550 nm pump is low; however, the lasing output may decrease or even be completely suppressed as the 980 nm pump is increased when the 1550 nm pump is strong (red solid line and black dotted line).  This behavior is similar to our experimental observations.   It should be noted that our model cannot explain the right ascending parts in plots $\rm \uppercase\expandafter{\romannumeral4}$ and $\rm \uppercase\expandafter{\romannumeral5}$ of Fig. \ref{fig2}(a). This phenomenon may originate from our simplification of the energy manifold to an energy level considering that the lasing has a blue shift tendency; in fact, the dynamics of the transitions between different energy sublevels of one manifold is not reflected in the absorption and emission cross-sections, thus a more complete model is needed to explain the phenomenon.

In order to further understand the lasing suppression phenomenon, we depict some parameters of the Yb:Er co-doped laser evolving with respect to the 980 pump in Fig. \ref{fig6}(d)-(f), corresponding to the black dotted line in Fig. \ref{fig6}(a). Figures \ref{fig6}(d) and (e) illustrate the normalized intracavity power and gain coefficient for the pump and laser modes, respectively, whereas Fig. \ref{fig6}(f) depicts the population percentage of erbium ions in different levels. The emergence of  lasing suppression can be intuitively explained as follows. When the 980 nm pump power is low, the gain provided for the 1600 nm lasing mode is mainly due to the 1550 nm pump (see Fig. \ref{fig5}(a) and (b)), thus the 1550 nm gain coefficient is negative, as shown in Fig. \ref{fig6}(e). The 980 nm pump excites the Er$^{3+}$ ions to the $\rm ^4I_{11/2}$ state, resulting in a reduction of the ground state population. Therefore, the output laser power increases with the 980 nm pump initially. As more and more ions are transferred to the $\rm ^4I_{11/2}$ level (Fig. \ref{fig5}(c)), the gain coefficient of the 1550 nm pump increases until, eventually, it is positive;  hence, the intracavity power of the 1550 nm pump keeps increasing, thereby decreasing the population in the metastable state, $\rm ^4I_{13/2}$, back to the ground state through stimulated emission. In other words, the 1550 nm pump field within the cavity opens a loss-path for the gain provided by the 980 nm pump. There is a competition between the 1550 nm field and the 1600 nm field \cite{liu2016gain}. Strictly speaking, the 1550 nm light should not be termed a \textit{pump} in this situation since its role has changed. 

For the pure Er$^{3+}$ laser case, the absorption cross-section of 980 nm is much smaller than for the co-doped case and it is comparable with that for the 1550 nm pump. Therefore, this situation is akin to the Yb:Er co-doped laser pumped by by a low power 980 nm source. In principle, lasing suppression should be observable as long as the 980 nm pump power is sufficiently high. However, in our experiments, this effect is limited by the maximum ouptut power available from our 980 nm laser (i.e., 76.5 mW).  In contrast, for the 980 nm and 1440 nm pumped Er:Yb co-doped laser, no such lasing suppression mechanism exists, since the ratio of the emission to absorption cross-sections is much lower than that for 1600 nm. The main function of the 1440 nm light is to transfer the Er$^{3+}$ ions from the ground state to the metastable state rather than inducing stimulated emission. While we did not perform a set of experiments using a 1480 nm pump, we expect that the results would be similar to those for the 1440 nm pump. The 980 nm and 1480 nm  co-pumping scheme has been already widely explored in Er$^{3+}$ doped amplifiers \cite{sini,delavaux1992hybrid,koch1997optical}.


\section{Summary}
In this work, we investigated the output behavior of a Yb:Er co-doped  microsphere laser when it was pumped with two light sources at different wavebands. We show that the two pumps may contribute both constructively and destructively to the lasing generation depending on their powers and wavelengths, and the lasing can be completely suppressed  when both of the pumps are strong. The observations are qualitatively explained by our simple theoretical model, which reveals that these phenomena originate from the diversity of energy levels in the gain medium. Aside  from its use for Yb:Er co-doped lasers, these phenomena should exist in lasers with many other gain media. Considering most earlier studies are based on a single pump, there may be many as of yet unrevealed phenomena in lasers that involve two or more pumps. In particular, it may be worth exploring whether such effects emerge when studying quantum behaviors of single-atom lasers \cite{mckeever2003experimental,astafiev2007single}. Moreover, the phenomenon of pump-induced lasing suppression  may lead to new methods for controlling lasing and may add new insights into laser operation.

\section*{ACKNOWLEDGMENTS}
This work was supported by the Okinawa Institute of Science and Technology Graduate University.


\begin{thebibliography}{10}
\newcommand{\enquote}[1]{``#1''}

\bibitem{koechner2013solid}
W.~Koechner, \emph{Solid-state laser engineering}, vol.~1 (Springer, 2013).

\bibitem{becker1999erbium}
P.~M. Becker, A.~A. Olsson, and J.~R. Simpson, \emph{Erbium-doped fiber
  amplifiers: fundamentals and technology} (Academic press, 1999).

\bibitem{sini}
K.~D. V.~Sinivasagam, M. A. G.~Abushagur and F.~Tumiran, \enquote{New pumping
  scheme for high gain and low noise figure in an erbium-doped fiber
  amplifier,} , IEICE Electron. Expresss \textbf{2}, 154--158 (2005).

\bibitem{Bray99}
M.~Bray, T.~Reid, K.~Jones, and M.~Poettcker, \enquote{Model to predict
  spectral shape of an hybrid 980nm \& 1480nm pumped erbium doped fibre
  amplifier,} in \enquote{Optical Fiber Communication Conference and the
  International Conference on Integrated Optics and Optical Fiber
  Communication,}  (Optical Society of America, 1999), p. WG4.

\bibitem{koch1997optical}
T.~L. Koch, \emph{Optical Fiber Telecommunications III}, vol.~2 (Academic
  Press, 1997).

\bibitem{choi2001high}
B.-H. Choi, H.-H. Park, M.~Chu, and S.~K. Kim, \enquote{High-gain coefficient
  long-wavelength-band erbium-doped fiber amplifier using 1530-nm band pump,}
  IEEE Photon. Technol. Lett. \textbf{13}, 109--111 (2001).

\bibitem{lee1999enhancement}
J.~Lee, U.-C. Ryu, S.~J. Ahn, and N.~Park, \enquote{Enhancement of power
  conversion efficiency for an {L}-band {EDFA} with a secondary pumping effect
  in the unpumped {EDF} section,} IEEE Photon. Technol. Lett. \textbf{11},
  42--44 (1999).

\bibitem{young2004efficient}
Y.~E. Young, S.~D. Setzler, K.~J. Snell, P.~A. Budni, T.~M. Pollak, and
  E.~Chicklis, \enquote{Efficient 1645-nm {E}r:{YAG} laser,} Opt. Lett.
  \textbf{29}, 1075--1077 (2004).

\bibitem{reynolds2017fluorescent}
T.~Reynolds, N.~Riesen, A.~Meldrum, X.~Fan, J.~M. Hall, T.~M. Monro, and
  A.~Fran{\c{c}}ois, \enquote{Fluorescent and lasing whispering gallery mode
  microresonators for sensing applications,} Laser Photon. Rev. \textbf{11},
  1600265 (2017).

\bibitem{he2013whispering}
L.~He, {\c{S}}.~K. {\"O}zdemir, and L.~Yang, \enquote{Whispering gallery
  microcavity lasers,} Laser Photon. Rev. \textbf{7}, 60--82 (2013).

\bibitem{he2011detecting}
L.~He, {\c{S}}.~K. {\"O}zdemir, J.~Zhu, W.~Kim, and L.~Yang, \enquote{Detecting
  single viruses and nanoparticles using whispering gallery microlasers,}
  Nature Nanotech. \textbf{6}, 428--432 (2011).

\bibitem{hodaei2014parity}
H.~Hodaei, M.-A. Miri, M.~Heinrich, D.~N. Christodoulides, and M.~Khajavikhan,
  \enquote{Parity-time--symmetric microring lasers,} Science \textbf{346},
  975--978 (2014).

\bibitem{feng2014single}
L.~Feng, Z.~J. Wong, R.-M. Ma, Y.~Wang, and X.~Zhang, \enquote{Single-mode
  laser by parity-time symmetry breaking,} Science \textbf{346}, 972--975
  (2014).

\bibitem{peng2014loss}
B.~Peng, {\c{S}}.~{\"O}zdemir, S.~Rotter, H.~Yilmaz, M.~Liertzer, F.~Monifi,
  C.~Bender, F.~Nori, and L.~Yang, \enquote{Loss-induced suppression and
  revival of lasing,} Science \textbf{346}, 328--332 (2014).

\bibitem{yang2016tunable}
Y.~Yang, F.~Lei, S.~Kasumie, L.~Xu, J.~Ward, L.~Yang, and S.~Nic~Chormaic,
  \enquote{Tunable erbium-doped microbubble laser fabricated by sol-gel
  coating,} Opt. Express \textbf{25}, 1308 (2017).

\bibitem{ward2016glass}
J.~M. Ward, Y.~Yang, and S.~Nic~Chormaic, \enquote{Glass-on-glass fabrication
  of bottle-shaped tunable microlasers and their applications,} Sci. Rep.
  \textbf{6}, 25152 (2016).

\bibitem{ward2008taper}
J.~M. Ward, P.~F{\'e}ron, and S.~Nic~Chormaic, \enquote{A taper-fused
  microspherical laser source,} IEEE Photon. Technol. Lett. \textbf{20},
  392--394 (2008).

\bibitem{wu2010ultralow}
Y.~Wu, J.~M. Ward, and S.~Nic~Chormaic, \enquote{Ultralow threshold green
  lasing and optical bistability in {ZBNA} ({Z}r{F} 4--{B}a{F}
  2--{N}a{F}--{A}l{F} 3) microspheres,} J. Appl. Phys. \textbf{107}, 033103
  (2010).

\bibitem{peng2016chiral}
B.~Peng, {\c{S}}.~K. {\"O}zdemir, M.~Liertzer, W.~Chen, J.~Kramer,
  H.~Y{\i}lmaz, J.~Wiersig, S.~Rotter, and L.~Yang, \enquote{Chiral modes and
  directional lasing at exceptional points,} PNAS \textbf{113}, 6845--6850
  (2016).

\bibitem{yang2005erbium}
L.~Yang, T.~Carmon, B.~Min, S.~M. Spillane, and K.~J. Vahala,
  \enquote{Erbium-doped and raman microlasers on a silicon chip fabricated by
  the sol--gel process,} Appl. Phys. Lett. \textbf{86}, 091114 (2005).

\bibitem{cai2002laser}
Z.~Cai, A.~Chardon, H.~Xu, P.~F{\'e}ron, and G.~M. St{\'e}phan, \enquote{Laser
  characteristics at 1535 nm and thermal effects of an er: Yb phosphate glass
  microchip pumped by ti: sapphire laser,} Opt. Commun. \textbf{203}, 301--313
  (2002).

\bibitem{arnaud2004microsphere}
C.~Arnaud, M.~Boustimi, M.~Brenci, P.~Feron, M.~Ferrari, G.~Nunzi-Conti,
  S.~Pelli, and G.~C. Righini, \enquote{Microsphere laser in $\rm
  {E}r^{3+}$-doped oxide glasses,} in \enquote{5th Iberoamerican Meeting on
  Optics and 8th Latin American Meeting on Optics, Lasers, and Their
  Applications,}  (International Society for Optics and Photonics, 2004), pp.
  315--320.

\bibitem{white2007resonant}
J.~O. White, M.~Dubinskii, L.~D. Merkle, I.~Kudryashov, and D.~Garbuzov,
  \enquote{Resonant pumping and upconversion in 1.6 $\mu$m $\rm {E}r^{3+}$
  lasers,} JOSA B \textbf{24}, 2454--2460 (2007).

\bibitem{ward2007optical}
J.~M. Ward, D.~G. O\'~Shea, B.~J. Shortt, and S.~Nic~Chormaic, \enquote{Optical
  bistability in er-yb codoped phosphate glass microspheres at room
  temperature,} J. Appl. Phys. \textbf{102}, 023104 (2007).

\bibitem{paschotta1997ytterbium}
R.~Paschotta, J.~Nilsson, A.~C. Tropper, and D.~C. Hanna,
  \enquote{Ytterbium-doped fiber amplifiers,} IEEE J. Quantum Electron.
  \textbf{33}, 1049--1056 (1997).

\bibitem{wu2003fluorescence}
R.~Wu, J.~D. Myers, M.~J. Myers, and C.~F. Rapp, \enquote{Fluorescence lifetime
  and 980nm pump energy transfer dynamics in erbium and ytterbium co-doped
  phosphate laser glasses,} in \enquote{High-Power Lasers and Applications,}
  (International Society for Optics and Photonics, 2003), pp. 11--17.

\bibitem{paschotta}
R.~Paschotta, \enquote{article on 'erbium-doped gain media' in the encyclopedia
  of laser physics and technology,}
  \url{https://www.rp-photonics.com/erbium_doped_gain_media.html}. Accessed on
  2017-05-08.

\bibitem{da2003mechanism}
L.~Da~Vila, L.~Gomes, L.~Tarelho, S.~Ribeiro, and Y.~Messadeq,
  \enquote{Mechanism of the {Y}b--{E}r energy transfer in fluorozirconate
  glass,} J. Appl. Phys. \textbf{93}, 3873--3880 (2003).

\bibitem{lei2014dynamic}
F.~Lei, B.~Peng, {\c{S}}.~K. {\"O}zdemir, G.~L. Long, and L.~Yang,
  \enquote{Dynamic {F}ano-like resonances in erbium-doped
  whispering-gallery-mode microresonators,} Appl. Phys. Lett. \textbf{105},
  101112 (2014).

\bibitem{liu2016gain}
X.-F. Liu, F.~Lei, M.~Gao, X.~Yang, C.~Wang, {\c{S}}.~K. {\"O}zdemir, L.~Yang,
  and G.-L. Long, \enquote{Gain competition induced mode evolution and
  resonance control in erbium-doped whispering-gallery microresonators,} Opt.
  Express \textbf{24}, 9550--9560 (2016).

\bibitem{delavaux1992hybrid}
J.-M. Delavaux, C.~Flores, R.~Tench, T.~Pleiss, T.~Cline, D.~DiGiovanni,
  J.~Federici, C.~Giles, H.~Presby, J.~Major \emph{et~al.}, \enquote{Hybrid
  {E}r-doped fibre amplifiers at 980-1480 nm for long distance optical
  communications,} Electron. Lett. \textbf{28}, 1642--1643 (1992).

\bibitem{mckeever2003experimental}
J.~McKeever, A.~Boca, A.~D. Boozer, J.~R. Buck, and H.~J. Kimble,
  \enquote{Experimental realization of a one-atom laser in the regime of strong
  coupling,} Nature \textbf{425}, 268--271 (2003).

\bibitem{astafiev2007single}
O.~Astafiev, K.~Inomata, A.~Niskanen, T.~Yamamoto, Y.~A. Pashkin, Y.~Nakamura,
  and J.~Tsai, \enquote{Single artificial-atom lasing,} Nature \textbf{449},
  588--590 (2007).

\end{thebibliography}
\end{document}